# METHANOL MASERS AS TRACERS OF CIRCUMSTELLAR DISKS

R. P. Norris[1], S. E. Byleveld[1,2], P. J. Diamond[3], S. P. Ellingsen[1,4], R. H. Ferris[1], R. G. Gough[1], M. J. Kesteven[1], P. M. McCulloch[4], C. J. Phillips[4], J. E. Reynolds[1], A. K. Tzioumis[1], Y. Takahashi[5], E. R. Troup[1], K. J. Wellington[1]

1) Australia Telescope National Facility, CSIRO, PO Box 76, Epping, 2121, Australia. (Email: rnorris@atnf.csiro.au)

2) School of Physics, University of Sydney, Sydney, 2006, Australia

3) National Radio Astronomy Observatory, PO Box 0, Socorro, NM 87801-0387, USA

4) School of Physics, University of Tasmania, GPO Box 252-21, Hobart, Tas 7001, Australia.

5) Kashima Space Research Centre, CRL, 893-1 Hirai, Kashima, Ibaraki, 314, Japan.

ABSTRACT

We show that in many methanol maser sources the masers are located in lines, with a velocity gradient along them which suggests that the masers are situated in edge-on circumstellar, or protoplanetary, disks. We present VLBI observations of the methanol maser source G309.92+0.48, in the 12.2 GHz $(2_0 - 3_{-1}E)$ transition, which confirm previous observations that the masers in this source lie along a line. We show that such sources are not only linear in space but, in many cases, also have a linear velocity gradient. We then model these and other data in both the 6.7 GHz $(5_1 - 6_0A^+)$ and the 12.2 GHz $(2_0 - 3_{-1}E)$ transition from a number of star formation regions, and show that the observed spatial and velocity distribution of methanol masers, and the derived Keplerian masses, are consistent with a circumstellar disk rotating around an OB star. We consider this and other hypotheses, and conclude that about half of these methanol masers are probably located in edge-on circumstellar disks around young stars. This is of particular significance for studies of circumstellar disks because of the detailed velocity information available from the masers.

**1) INTRODUCTION**

A fresh wave of studies of masers as probes of star formation has been stimulated by the discovery of the strong methanol $(2_0 - 3_{-1}E)$ maser transition at 12.2 GHz by Batrla et al. (1987), and then the discovery of the strong methanol $(5_1 - 6_0A^+)$ transition at 6.7 GHz by Menten (1991). Although methanol masers at other transitions have also been extensively studied (see for example Elitzur 1992, for a review), the 6.7 and 12 GHz transitions are particularly strong and accessible to synthesis instruments, and so here we restrict the discussion to these two transitions. Extensive surveys (e.g. Norris et al. 1987; Koo et al. 1988; Kemball, Gaylard, & Nicolson 1988; MacLeod & Gaylard 1992; MacLeod, Gaylard, & Nicolson 1992; MacLeod, Gaylard, & Kemball 1993; Gaylard & MacLeod 1993; Caswell et al. 1993, 1995; Schutte et al. 1993; Ellingsen et al. 1996a) have shown these methanol masers to be extremely common in Galactic star formation regions, but none is known outside our Galaxy (Ellingsen et al. 1994a) with the exception of two masers in the LMC (Sinclair et al. 1992; Ellingsen et al. 1994b). These surveys also show that they are frequently associated with OH and H$_2$O masers, which in turn are known to be generally associated with compact HII regions. Thus we surmise that most of the methanol masers are associated with compact HII regions, powered by young O or B stars.

The current interest in methanol masers arises not only because they provide information additional to that provided by OH and H$_2$O, but also because the structures of some methanol masers are much simpler than those of OH and H$_2$O masers. The first maps of methanol sources were made with a single-baseline interferometer by Norris et al. (1988; subsequently N88), who commented that some of the sources showed linear structures that had not been seen before in OH or H$_2$O masers. The first maps at 6.7 GHz by Norris et al. (1993; subsequently N93) showed that this effect was relatively widespread. Only a few sources have so far been mapped by interferometers or synthesis instruments (N88, N93; Menten et al. 1988a,b; Menten et al. 1992), but about half of those show simple linear structures with an approximate velocity gradient along them. This surprising result is in contrast to OH and H$_2$O masers, where no maser in a star formation region has ever been observed to have the tightly collimated linear structure that seems relatively common in methanol masers, although a few OH and H$_2$O sources do have a roughly elongated morphology (e.g. Norris et al. 1982), and linear structures are seen in high-velocity masers (e.g. te Lintel Hekkert et al. 1995). For convenience in this paper, we refer to those masers that are located along straight or curved lines as "linear", and those that have a complex or very compact morphology as "complex".

Unfortunately, most of the methanol sources that have been mapped lie in the Southern Hemisphere, while most of the OH and H$_2$O masers and compact HII regions that have been mapped with high resolution lie in the Northern Hemisphere. There are thus very few sources where high-resolution information exists both on the methanol masers and at other wavelengths in the same source. In only one case (Ellingsen et al. 1996b) do we know how a linear maser source is related to the parent HII region. We consider it significant that in this case the masers lie across a diameter of the HII region.

N93 suggested that the lines of masers may represent circumstellar disks (also referred to as accretion disks and protoplanetary disks in the literature) seen edge-on, because the spatial scales and velocity gradients are approximately those expected for such disks. Circumstellar disks appear to be very common around young stars and contain large amounts of warm molecular gas (e.g. Beckwith & Sargent 1993), and so appear to be likely sites for molecular masers. The principal objective of this paper is to examine the evidence for the hypothesis that the strong methanol masers lie within these circumstellar disks.

In Section 2 of this paper, we present details of very long baseline interferometry (VLBI) observations of the $2_0 - 3_{-1}E$ (12 GHz) transition of methanol in the source G309.92+0.48. In Section 3 we present the results of these observations, and compare them to the earlier 6.7 and 12.2 GHz maps. In Section 4 we consider the morphology of methanol masers, and in Section 5 we discuss the alternative hypotheses that might produce the observed linear features, and model our results in terms of a rotating circumstellar disk.

## 2) OBSERVATIONS AND DATA REDUCTION

G309.92+0.48 was observed in three separate VLBI experiments in 1989 November, 1990 April and July, with the 64 m Parkes antenna of the Australia Telescope National Facility (ATNF), the 70 m Tidbinbilla antenna of the NASA Deep Space Network, and the Mount Pleasant 26 m antenna of the University of Tasmania. The data were recorded on MkII VLBI tapes and then correlated on the MkII correlator operated by the National Radio Astronomy Observatory (NRAO). The individual antennas were equipped with room-temperature 12 GHz receivers, giving system temperatures ~ 250 K.

We observed G309.92+0.48 for a total of 10 hr over a wide range of hour angle, together with calibration observations of extragalactic continuum sources. Observations from the separate VLBI experiments were combined after checking them for consistency of calibration. We used a bandwidth of 0.5 MHz, divided into 96 frequency channels at correlation, to give a velocity resolution of 0.3 km s$^{-1}$ after Hanning smoothing. Because of instrumental limitations at the time, we observed only one linear polarization, but since the masers are not strongly linearly polarized (Caswell et al. 1993), and because any small effects would be averaged out by the range of parallactic angles, this should not produce any significant errors. Before combining the different epochs of data, we checked the individual data sets for consistency, and were unable to detect any changes in amplitude or component separation between epochs.

All the data were calibrated and reduced using the AIPS processing system. We calibrated the bandpasses using additional observations of the continuum source 3C 273, and calibrated the amplitudes by normalizing the cross-correlation spectra by the simultaneous auto-correlation spectra from each antenna.

We then selected one spectral feature (the -59.80 km s$^{-1}$ feature - subsequently referred to as the reference feature) which, since its amplitude was constant with hour angle, appeared to be unresolved. We used this unresolved reference feature to calibrate the phase of the data on other features in the source, by subtracting its phase from the phase of every other spectral channel. We then mapped and CLEANed the data to produce an image of the maser spots corresponding to the emission in each spectral feature.

The size (FWHM) of the synthesized beam of the Australian VLBI Array at 12.2 GHz in this experiment was 0.011 * 0.007" which is comparable to, or larger than, the size of the individual maser spots (McCutcheon et al. 1988; Menten et al. 1992; Ellingsen et al. 1996b). However, our relatively poor uv coverage, together with limited amplitude calibration information meant that, in this experiment, our dynamic range in each channel map was in the range 10-30, and we could not measure accurate maser spot sizes. All observed maser spots were strongly visible on all baselines, with no indication of resolution effects, except that component 'e' is weaker in the VLBI data than in single-dish spectra. Our data are not of sufficient quality to determine unambiguously whether this is due to resolution or to other instrumental effects. Otherwise, within our estimated 20% amplitude calibration uncertainties, we believe that essentially all the flux appearing on the spectra is represented in the maps.

The positional errors on VLBI maps are dominated not by thermal noise but by systematic effects, and are notoriously difficult to estimate. As discussed below, comparison of these results with earlier PTI positions implies that the errors are of order 5 milliarcsec in most cases.

### 3) THE VLBI RESULTS

The resulting map and spectrum are shown in Figure 1, and the measured relative positions of the individual spots are shown in Table 1. These VLBI positions agree with the Parkes-Tidbinbilla Interferometer (PTI) positions obtained by N88 to within 5 milliarcsec, except for feature 'e', which shows a difference in position of 22 milliarcsec. Thus the new VLBI map essentially confirms the earlier lower resolution maps, except that four new features (x1 to x4) have been identified in the VLBI data. These features were not visible in the earlier data because of the lower quality of those data. However, we note that these new features follow the same alignment as the features already known.

It is significant that the new VLBI maps, containing four new maser features, do not show any structure other than the line of features revealed in the Australia Telescope Compact Array (ATCA) and PTI data. This confirms that ATCA maps are reliable for determining the overall structure of these maser sources, and that the technically more difficult VLBI maps are needed only when the detailed fine-scale structure or accurate positions are needed, as, for example, when measuring the proper motions of the masers.

### 4) THE MORPHOLOGY OF METHANOL MASER SOURCES

In Table 2 we list all the methanol sources for which interferometer or synthesis images are available. Of the 17 methanol maser sources listed, nine show a linear or curved morphology. We refer to those masers as "linear". The remaining eight are either very compact or show a more complex distribution that is similar to the complex distributions seen in OH and $H_2O$ masers, and those we refer to as "complex". In none of the linear sources studied here have the OH and $H_2O$ masers been studied with sufficient resolution to make a detailed comparison with methanol, and none of the northern OH and $H_2O$ masers which have been studied in detail has detailed maps of the methanol masers, with the exception of W3(OH) (Menten et al. 1988a,b) which is complex. We therefore have no information at present on the distribution of the OH and $H_2O$ masers in linear methanol maser sources. However, we assume that the OH masers found associated with the linear methanol sources will be as complex as the northern ones that <u>have</u> been studied. These OH and $H_2O$ masers generally appear to lie outside their parent HII region, and so we assume that this is also the case for the linear methanol sources.

The remaining discussion is confined to the nine linear sources, together with one source (G351.42+0.64) which, on further examination, may be two linear sources. Figures 2 to 11 show the spectra and maps of the masers for each source, together with a *v-a* diagram (discussed below) for each. In each case, we have taken the 6.7 GHz data from N93, and combined it with the 12 GHz data from N88 where available. Because accurate absolute positions are not available for the 12 GHz masers, we have combined the data by moving the 12 GHz masers relative to the 6 GHz masers such that features common to both transitions are coincident. In some cases (e.g. G345.01+1.79) there are so many features in common, with so little positional error, that the identification is beyond doubt. In other cases the identification is less certain, and we have identified those pairs of features at each transition whose velocities and positional separations are most closely matched. We then determined the major axis of the combined cluster using a least-squares fit, and then measured the separations *a* of each maser spot when projected onto the major axis of the cluster, and plotted that against its velocity *v*. It is remarkable that several of the sources show a line not only in the R.A. - Decl. plane, but also in the *v-a* plane.

## 5) DISCUSSION
### 5.1) THE ORIGIN OF LINEAR METHANOL MASER SOURCES

N88 noted that several of the masers they mapped showed a linear or arc-like morphology, and this was confirmed by the 6.7 GHz observations of N93. This effect has now been observed in too many sources and using too many different techniques to be dismissed as an artefact of either chance alignments or technical problems. Of the 17 sources (listed in Table 2) now mapped by interferometers or synthesis instruments, nine (shown in Figures 2-11) may be described as essentially linear or curvilinear. We now consider three possible ways in which a linear morphology might arise in a star formation region.

*5.1.1) Jets and collimated outflows.*

A linear morphology could, in principle, indicate a jet emanating from the star. Outflows of gas with velocities of a few km s$^{-1}$ are common within star formation regions (e.g. Bachiller & Gómez-González 1992) , although they are broad and uncollimated, and quite unlike the tightly collimated lines we see in the methanol masers. On the other hand, narrow collimated jets (e.g. Rodriguez et al. 1986,1989 1990; Torelles et al. 1989; Mundt, Brugel, & Bührke 1987; te Lintel Hekkert et al. 1995) have been seen in other star formation regions, but these are associated with high-velocity flows.

These high velocities contrast sharply with the low velocities seen in the methanol masers. Furthermore, the high energy processes needed to produce and collimate these jets are unlikely to leave the fragile methanol molecules intact. Thus, there is no observational or theoretical support for the methanol lines to represent highly collimated outflows or jets. We note that a specific prediction of the jet hypothesis is that the line of masers should be radial to the parent star or HII region.

*5.1.2) Shock fronts*

A layer of cool dense dust and gas is expected to lie between the shock and ionization fronts surrounding a compact HII region, and it has been suggested that OH masers in particular lie within this zone (e.g. Elitzur 1992). In the case of the complex source W3(OH), Menten et al. (1988b) have shown that the OH and methanol masers occupy roughly the same region. For this to be true for the linear methanol sources too, it is necessary to provide a mechanism that will produce the observed narrow lines of methanol masers. For example,  if methanol masers were to lie close to a very smooth regular shock front, then they could appear as lines or arcs if observed edge-on. A spherically symmetric HII region would then be seen as a circle of masers surrounding the HII region, although we might expect inhomogeneities or anisotropies to break up this circle to produce an arc similar to that seen in Figure 1. However, we have no evidence that such smooth regular shock fronts exist, since in all observed cases, such as W3(OH), there seems to be considerable irregularity.

This hypothesis also suffers from the problem that it does not explain the observed smooth velocity gradients in any natural way. It also predicts that the lines of masers should occur around the circumference of the compact HII region.

### *5.1.3) Edge-on circumstellar disks*

Circumstellar disks have been observed around a number of young stars using a variety of techniques (e.g. Smith & Terrile 1984; Telesco et al. 1988; O'Dell & Wen 1994), and are also expected to form around massive stars (Hollenbach et al. 1994). Such disks are now believed to be common around young stars and contain large amounts of warm molecular gas (e.g. Beckwith & Sargent 1993). There is thus a strong *prima facie* case to expect molecular masers to occur within these disks. If the methanol masers occur in these disks, and these disks are observed edge on, then this would naturally explain the observed linear structures, and would predict that these masers lie across a diameter of the HII region. If some disks were observed at a slightly inclined angle, with the masers occurring in a limited range of radii within the disk, then this may also explain the curved arcs we observe in sources such as G309.92+0.48. However, the large fraction of masers with linear structure implies that a disproportionate number of these disks are edge-on, and in Section 5.3 below we discuss possible reasons for this.

### **5.2) Comparison with continuum data**

Ellingsen et al. (1996b) have imaged the compact continuum sources associated with the methanol maser sources G339.88-1.2 and G351.42+0.64. The continuum source in G351.42+0.64 is, like the methanol masers in that source, complex, and may be the result of several energizing stars. It is therefore difficult to interpret this source in terms of any simple model. In G339.88-1.26, on the other hand, there is a simple compact continuum source with a slight extension to the north-east, suggesting an outflow along this axis. The line of masers in this source lies across a diameter of the HII region, and is perpendicular to the continuum extension. This result and orientation is consistent with the edge-on circumstellar disk hypothesis, and appears to be inconsistent with the jet hypothesis, unless we invoke two jets emitted symmetrically from the star, perpendicularly to the continuum extension. The observation of G339.88-1.26 is also inconsistent with the shock-front hypothesis, which predicts that the masers surround the HII region. Although this result appears to support the circumstellar disk hypothesis, we acknowledge that only one linear source has so far been observed, and that observations of other sources are needed before this particular evidence can be considered conclusive.

In G339.88-1.26, both the jet and the shock hypotheses fail to explain the velocity gradient of a few km s$^{-1}$ along the length of the maser line, which is explained by the circumstellar disk hypothesis in a natural way. We therefore favour the circumstellar disk hypothesis. However, an arbitrary distribution of the disks' inclination angles is inconsistent with the large fraction of disks that we observe edge-on, and so our model must explain why we see so many edge-on disks.

**5.3) Why do we see so many edge-on disks?**

We propose two mechanisms that will produce selection effects which make us see more edge-on disks than disks at other inclinations.

The first mechanism (mechanism a), shown in Figure 12(a), concerns the conditions under which maser action occurs. In an interstellar gas cloud containing excited molecules, maser action will occur predominantly along those columns through which there is a maximum number of molecules with the same line-of-sight velocity (so that they radiate at the same frequency). In a disk, much greater column depths are found in maser columns lying within the plane of the disk than those inclined to that plane. Other effects caused by differential rotation are discussed below in Section 5.4.

Assuming an $H_2$ density of $10^7$ cm$^{-3}$, a methanol/$H_2$ abundance of $10^{-7}$ (Menten et al. 1988c), and a column density of $3.10^{16}$ cm$^{-2}$ (Cragg et al. 1992), we derive a column length of $3.10^{16}$ cm. This enormous length is probably greater than is physically realistic, indicating that the column density or another parameter has been overestimated. However, even after reducing it by an order of magnitude, it is still comparable with the radius of the disk (Hollenbach et al. 1994), and therefore can only be achieved by maser columns lying in the plane of the disk. Therefore the masers, which are strongly beamed, will radiate more strongly in the plane of the disk than perpendicular to that plane. Thus the strongest masers observed will be those in edge-on disks.

The second mechanism (mechanism b), shown in Figure 12(b), concerns the optical depth of the HII region surrounding the parent star. If the parent OB star is surrounded by a disk, then that disk will bisect the compact HII region. The high density of neutral material in the disk shields the fragile molecules within it from the ultraviolet photons in the surrounding HII region, so the interior of the disk remains un-ionized. Masers within the disk that radiate in the plane of the disk will beam their emission through this neutral material into the interstellar medium. However, masers within the disk that radiate perpendicular to the disk will beam their emission through the HII region. Hollenbach et al. (1994) have shown that this HII region is likely to be optically thick at centimetre wavelengths, and so the maser emission will be absorbed. Thus, once again, the strongest masers observed will be those in edge-on disks.

It is not clear which of these two mechanisms will dominate. In mechanism b, the optical depth is proportional to $\nu^{-2.1}$, and so will be less effective for the 12 GHz transition than for the 6 GHz transition. Therefore, if this mechanism dominates, we expect to see fewer linear maser sources at 12 GHz than at 6 GHz. We see no indication of this in our data, leading us to favour mechanism a.

Both mechanisms predict that the strongest maser sources should be in edge-on disks, but weaker maser sources may be found in disks at other angles, so that in any flux-limited sample, the sample will contain a disproportionately large number of edge-on disks. However, if the required maser column length is sufficiently long (mechanism a) or the optical depth of the HII region is sufficiently high (mechanism b) then perhaps all methanol masers are necessarily in edge-on disks. We see no evidence for any correlation between morphology and peak flux in our small sample of sources, which may indicate that the effect is dominant even for the weakest masers studied here.

**5.4 The geometry of the maser spots**

Because we observe the masers in projection, we have no information on the physical location of the masers in the disk. Maser action occurs along those lines of sight where there is a sufficient optical depth of excited gas at a similar velocity. In a uniform Keplerian disk with no internal motion, this coherent optical depth reaches a maximum tangentially at the limbs and radially in front of the central mass, giving a double- or triple-peaked spectral profile, depending on the thickness of the masing region of the disk (Ponomarev, Smith, & Strelnitski 1994; Watson & Wallin 1994). Such a triple spectrum is demonstrated by the spectacular Keplerian water masers around the nucleus of NGC 4258 (Miyoshi et al. 1995).

However, the spectra of the methanol masers are clearly not double- or triple-peaked, and we attribute this to local velocity and density perturbations within the disk. The shape of the maser spectrum from a Keplerian disk depends not only on the geometry of the disk, but also on the internal gas motion within the disk. In the case of the circumnuclear disk in NGC 4258, its rotational velocity ($v_{rot} \sim 1000$ km s$^{-1}$) is much larger than the internal motion, and so the masers have a roughly triple-peaked profile, corresponding to three distinct clusters of masers. In the methanol maser disks, on the other hand, the rotational velocity of the disk ($v_{rot} \sim 4$ km s$^{-1}$) is comparable to the internal motion $v_{turb}$ of the gas, and so the positions at which maximum coherence length is achieved will be those at which regions of gas with similar velocities happen to be aligned along the line of sight. Such columns of maser emission need not be contiguous, but may involve widely separated regions of excited gas. However, the velocities of the masers will still cluster around the Keplerian rotation curve, with a velocity dispersion corresponding to the magnitude of the local velocity variations within the disk. The locations, and velocities, of the masers will also be affected by local density enhancement, and we may speculate that some of the masers may indicate regions of high density, or protoplanets, within the circumstellar disk.

We note that the ratio $v_{rot}/v_{turb}$ is critical in determining the maser spectrum, the observed spatial distribution of maser spots, and the observed *v-a* diagram. For large $v_{rot}/v_{turb}$, the spectrum of an edge-on disk will resemble NGC 4258: it will have a two- or three-peaked spectrum, and the maser spots will follow a Keplerian curve or else be confined to a single radius. For small $v_{rot}/v_{turb}$, the spectrum will be complex, the maser spots will still be located along a line in the spatial domain, but they will occupy all four quadrants of the *v-a* diagram. For $v_{rot}/v_{turb} \sim 1$, we expect a complex spectrum, but in the *v-a* diagram there will be some evidence of a velocity gradient, which will be expressed by the spots tending to fall into only two quadrants of the diagram. This latter case is most similar to the case that we observe in the methanol masers, from which we deduce that in most of these sources the internal velocity of the gas is similar to the rotational velocity of a few kms$^{-1}$, although in one or two cases, such as G345.01+1.79(S), the points in the *v-a* diagram fall along a line indication that $v_{rot} > v_{turb}$. We also note that G345.01+1.79(S) has one of the highest velocity ranges (presumably corresponding to $v_{rot}$ ) of the sources studied here, although the number of sources studied here is too small for us to determine whether or not this is significant.

Our knowledge of the details the methanol masers is poor, so it is difficult to estimate the length of the maser column, and hence the beaming angle. If it were possible to estimate the fraction of compact HII regions with linear methanol maser sources, then, by modelling the disks, we would be able to estimate the beaming angle of the masers. However, our statistics are not yet adequate to justify such a calculation, which moreover would be confused by the selection effects described in Section 5.3. We conclude that we know very little about the shape and position of the masers themselves, and instead confine ourselves in the rest of this paper to discussing the kinematics of the medium in which they are embedded.

## 5.5 Modelling the Disks

Modelling circumstellar disks (e.g. Ruden 1994) shows that the motions within the disk are expected to be essentially Keplerian, although the disks around higher-mass stars may contain a significant fraction of the mass of the star (Hollenbach et al. 1994) and so may be slightly non-Keplerian. Thus our circumstellar disk hypothesis implies that (a) the maser velocities should be roughly consistent with Keplerian rotation, and (b) modelling this rotation should result in a mass of the central star in the range 3-120 $M_\odot$ for an OB star. The jet and shock hypotheses make no such prediction. In this section we model the velocities of the masers.

Our knowledge of the conditions within the disk, and our knowledge of the maser pumping mechanism, are both poor. Thus we know neither the minimum radius from the parent star at which methanol can survive in the disk, nor the maximum radius at which there is sufficient radiation to pump the masers. Furthermore, the maser pumping mechanism itself may confine the maser emission to a limited range of radii, as happens in OH/IR stars (e.g. Elitzur 1992). In the following discussion, we consider the minimum radius $r_{min}$ and maximum radius $r_{max}$ at which maser emission can occur.

If $r_{min}$ and $r_{max}$ are similar, so that the masers are located in a thin ring around the star, then we expect the line-of-sight velocities $v(a)$ of the masers in an edge-on disk to be a linear function of observed distance $a$ from the star. Four of the sources (G305.21+0.21, G309.92+0.48, G336.43-0.26, G345.01+1.79(S)) do indeed show such a line in the $v$-$a$ diagram, and one complex source (G351.42+0.64) shows two.

The gradient of this line is a function $\frac{dv}{da} = \sqrt{G\,M/r_{max}^3}$ of the mass $M$ of the star and the radius $r_{max}$ of the ring. However, although the gradient $dv/da$ is well determined by the data, this gradient does not itself provide any useful physical parameters. Instead, the mass of the star is proportional to $r_{max}^3$. However, $r_{max}$ is poorly determined observationally, as the masers are not uniformly distributed around the ring, and the measurement of $a_{max}$, the observed extent of the masers, gives only a lower limit to $r_{max}$. Therefore, only a rough estimate of the stellar mass is obtained. For the masers whose *v-a* relationship is linear, these estimates of the ring size and stellar mass are given in Table 2. It is remarkable that all derived masses lie roughly in the range 3-120 $M_\odot$ appropriate to an OB star.

If $r_{min}$ and $r_{max}$ are well separated, then each ring of maser emission will produce a line whose gradient *dv/da* is a function of *r*. The ensemble of lines from each element of the disk will therefore produce a distribution of points which lies in only two quadrants of the *v-a* diagram, as shown in Figure 13. In this case $a_{max}$ again provides a lower limit to $r_{max}$ but the mass of the star now depends on an envelope to the quadrant, and so in this case the mass cannot be determined to any useful accuracy. However, we point out that the requirement for the masers to lie in only two quadrants of the *v-a* diagram is a sensitive test of the circumstellar disk hypothesis.

If the disk is not edge on, then the same result is obtained if we project the separation onto the major axis of the observed distribution. Thus the circumstellar disk hypothesis predicts that in all cases where we see the disk edge on, or nearly edge on, the maser points should lie in only two quadrants of the *v-a* diagram. This result is not specific to Keplerian rotation, but holds for any rotating disk.

Unfortunately, existing observations cannot constrain the location of the star relative to the line of masers with sufficient accuracy to be useful, and we have no information on the velocity of the central star. Thus our requirement that the masers reside in only two quadrants of a diagram with specified axes is weakened to the condition that we should be able to draw axes such that the masers reside in only two quadrants. This is still a sensitive test provided we demand that a significant number of maser components resides in both halves of the diagram.

In Figures 2-11 we have drawn axes on each *v-a* diagram, other than those five where the masers clearly lie along a line, such that the masers occur in only two quadrants. In each case it is indeed possible to do so, within uncertainties imposed by the data, with the exception of only one 12 GHz maser component in G339.88-1.26. We consider that this one spurious point, out of ~100 maser positions plotted, does not detract significantly from the otherwise impressive fit of the data to the model. The anomaly can probably be attributed to peculiar local motions, as discussed above in Section 5.4. In the case of G351.42+0.64, the *v-a* diagram shows two clear lines, indicating two separate centres of activity in this source. We note that the two lines show opposite senses of rotation, and that both senses are found equally in the sources discussed here. In view of the four sources which show a linear velocity gradient (indicating a ring, as discussed above), and the remaining sources (exemplified by G339.88-1.26) which show masers in only two quadrants of the *v-a* diagram, we consider it extremely unlikely that these alignments occur by chance.

In Table 2 we list the lower limit masses derived from these plots, as well as the diameter of the inferred circumstellar disk, as discussed above. Both these quantities depend critically on the extent of the masers, rather than being constrained by the gradient of the *v-a* correlation, and so are subject to large uncertainties.

In Figure 1(c) we show the *v-a* diagram of the new VLBI data on G309.92+0.48. Not only is the new plot consistent with that derived from the earlier data, but the four new points fit the same model. It is also noteworthy that more accurate positions from the VLBI data show a better fit to the rotational disk model than do the earlier, poorer quality, data. We consider this strong support for the circumstellar disk hypothesis.

**5.6) Complex sources and the relationship to other masers**

In this paper we have explicitly ignored the complex sources. It is possible that these are face-on disks, although they do not tend to be any weaker than the linear sources and so we consider this unlikely. In some cases, such as G351.42, they may be caused by the confusion of two adjacent sources. However, in the one complex source where detailed information is available, W3(OH) (Menten et al. 1988a,b), the methanol masers are roughly coincident with the OH masers. We know this cannot be the case for the linear masers, as linear arrangements of OH masers are unknown, and so it is possible that the complex methanol maser sources are intrinsically different from the linear sources.

In the linear sources the methanol masers probably lie in a circumstellar disk, while the OH and H$_2$O masers lie in the cool gas outside the HII region, between the shock and ionization fronts. In the case of the complex sources, the methanol and OH may co-exist in that same region.

We also note that the linear sources themselves appear to be further subdivided into two distinct classes - those which show a linear *v-a* plot (e.g. G309.92+0.48 and G345.01+1.79(S)) and those whose *v-a* plot shows two clusters of masers in opposite quadrants of the diagram (e.g. G339.88-1.26). While this appears to be a genuine division, the small number of sources so far studied does not yet permit us to study this effect in any further detail.

**5.7) Further observational tests**

We list two further observational tests of the circumstellar disk hypothesis.

1. Regardless of the kinematics of the masers, if we assume that they have a transverse velocity range similar to the observed radial velocity range, then a proper motion of order 0.0005" yr$^{-1}$ should be observed. Therefore, if the masers are contained within a rotating disk, we should be able to observe the orbital proper motion within a few years. We have already taken first-epoch VLBI observations for this test, and will continue to do so annually. There should also be a corresponding acceleration of the maser features within the spectrum (of ~0.01 km s$^{-1}$ yr$^{-1}$), but a longer time will be required before this effect is observable, as small changes may be masked by intensity variations within the masers.

2. If the predominance of linear sources is caused by the selection effects described above, then linear structures may be less common in weaker sources. We have started to explore this by mapping a larger sample of sources.

## 6) CONCLUSION

We have examined the evidence for the hypothesis that some methanol masers are contained within rotating edge-on circumstellar disks, and find that the data strongly support this hypothesis. New VLBI observations of G309.92+0.48 confirm the linear structure seen in earlier, lower resolution images of this source. Detailed modelling of this and nine other methanol maser sources with linear structure has shown that the velocities and positions of the masers in these sources are consistent with a model in which the masers are located in a rotating disk surrounding the parent star. Of the ~100 maser positions plotted, only one fails to support this model. The model also produces disk sizes and central masses which are consistent with theoretical models of circumstellar disks around OB stars. The predominance of edge-on sources is caused by one or both of two selection effects which cause edge-on sources to be more easily visible than face-on sources, so that about half the observed methanol maser sources are in edge-on disks. We conclude that methanol maser sources with linear morphology probably delineate rotating, edge-on, circumstellar disks.

## ACKNOWLEDGMENTS

We thank the Director and staff of the Tidbinbilla DSN station for their assistance with the VLBI observations. We also thank Peter te Lintel Hekkert and Jim Caswell for some helpful comments on an early draft of this paper.

| Feature | velocity (kms$^{-1}$) | RA (milliarcsec) | Dec (milliarcsec) | Flux (Jy) |
|---|---:|---:|---:|---:|
| a | -58.00 | 141.7 | 34.0 | 15.3 |
| x1 | -58.13 | 132.0 | 54.4 | 2.7 |
| x2 | -58.65 | 82.4 | 34.6 | 3.2 |
| b | -58.84 | 57.0 | 21.5 | 12.1 |
| x3 | -59.29 | 25.8 | 3.5 | 2.2 |
| c | -59.51 | 18.6 | 5.2 | 25.0 |
| x4 | -59.67 | 13.2 | 4.6 | 10.8 |
| d | -59.80 | 0.0 | 0.0 | 140.0 |
| e | -60.38 | -157.0 | -172.3 | 4.0 |

Table 1: Positions of the measured features in G309.92+0.48. Labels a-e correspond to the labels used by N93, and x1 to x4 are new features. The last column gives the measured cross-correlated flux density.

| Source | Refs | Kinematic distance (kpc) | Morphology | Diameter (AU) | Keplerian mass ($M_\odot$) |
|---|---|---|---|---|---|
| G305.21+0.21 | N93 | 8.0 | line | 1600 | 1 |
| G309.92+0.48 | N88, N93 | 5.5 | curve | 4300 | 10 |
| G318.95-0.20 | N93 | 2.0 | curve | 1000 | 4 |
| G323.74-0.26 | N88, N93 | 3.0 | complex | | |
| G328.23-0.53 | N93 | 2.9 | 2 clusters | | |
| G328.81+0.63 | N93 | 3.1 | line | 7300 | |
| G331.28-0.19 | N88, N93 | 4.8 | line | 1100 | |
| G336.43-0.26 | N93 | 6.7 | line | 2600 | 14 |
| G339.88-1.26 | N88, N93 | 3.0 | line | 3100 | |
| G340.78-0.10 | N93 | 9.4 | complex | | |
| G345.01+1.79(N) | N88, N93 | 2.3 | line | 600 | |
| G345.01+1.79(S) | N88, N93 | 2.3 | line | 670 | 4 |
| G351.42+0.64 | N88, N93 | 1.7 | 2 clusters | 3400,1200 | 9,2 |
| G351.78-0.54 | N93 | 2.2 | compact | | |
| G9.62+0.19 | N93 | 2.0 | complex | | |
| G188.95+0.89 | N88, N93 | ? | complex | | |
| W3(OH) | M88, M92 | 3.0 | complex | | |

Table 2: A list of the all the methanol sources for which interferometer or synthesis images are available. Diameters and masses are those derived from the modelling in this section. Diameters are given for linear sources only. Masses are given only for sources with a linear *dv/da* relationship, and are very uncertain. References are: N88: Norris et al. 1988; N93: Norris et al. 1993; M88: Menten et al. 1988a,b; M92: Menten et al. 1992.

FIGURE CAPTIONS

Fig 1. Single-dish spectrum (a),VLBI map (b), and *v-a* diagram (c) of G309.92+0.48. Note that not only does the map (b) show a linear structure, but a linear structure is also present in the *v-a* diagram, which is consistent with an edge-on rotating ring of masers. 12 and 6.7 GHz spectra are represented by broken and full lines respectively. The flux scale to the left of the spectrum refers to the 6.7 GHz data, and that on the right refers to the 12.2 GHz data. The origin of the velocity axis in (c) is arbitrary.

Fig. 2 Spectrum, map, and v-a diagram of G309.92+0.48. In this and subsequent diagrams: open circles represent 12 GHz masers from N88, and filled circles represent 6.7 GHz masers from N93; 12 and 6.7 GHz spectra are represented by broken and full lines respectively; the flux scale to the left of the spectrum refers to the 6.7 GHz data, and that on the right refers to the 12.2 GHz data; the cross in (b) represents the position angle and origin of the fitted major axis on to which the positions have been projected to obtain the position offsets shown in (c), and the origin of both axes in (c) are arbitrary.

Fig 3. Spectrum, map, and *v-a* diagram of G305.21+0.21.

Fig 4. Spectrum, map, and *v-a* diagram of G318.95-0.20. The dotted lines in (c), in this and subsequent figures, indicate the position and velocity of the parent star, such that all masers lie in only two quadrants, which is consistent with an edge-on rotating circumstellar disk.

Fig 5. Spectrum, map, and *v-a* diagram of G328.81+0.63

Fig 6. Spectrum, map, and *v-a* diagram of G331.28-0.19 The one maser feature in the "wrong" quadrant is consistent with the positional error estimated by N93 of 0.02 arcsec.

Fig 7. Spectrum, map, and *v-a* diagram of G336.43-0.26

Fig 8. Spectrum, map, and *v-a* diagram of G339.88-1.26. The 12 GHz component in the lower right-hand quadrant in (c) (corresponding to "g" in N88 and N93) is anomalous.

Fig 9. Spectrum, map, and *v-a* diagram of G345.01+1.79(N)

Fig 10. Spectrum, map, and *v-a* diagram of G345.01+1.79(S)

Fig 11. Spectrum, map, and *v-a* diagram of G351.42+0.64. The two clusters in the map (b) clearly appear as two lines in the *v-a* plane, indicating two separate rotating circumstellar disks around separate parent stars, with opposite senses of rotation.

Fig. 12. Diagram showing the two mechanisms which lead us to view circumstellar disks preferentially edge-on. In (a) the masers within the plane of the disk have a greater column length than those perpendicular to it, and so are brighter. In (b), any maser emission emitted perpendicular to the disk is absorbed by the hemispherical optically thick HII regions above and below the plane, whereas masers that emit within the plane suffer no absorption.

Fig 13. The rotation curves of a rotating disk. Each broken ring in the upper part of the diagram corresponds to a line in the lower part. The shaded area indicates the locus in the *v-a* diagram where masers might be found.

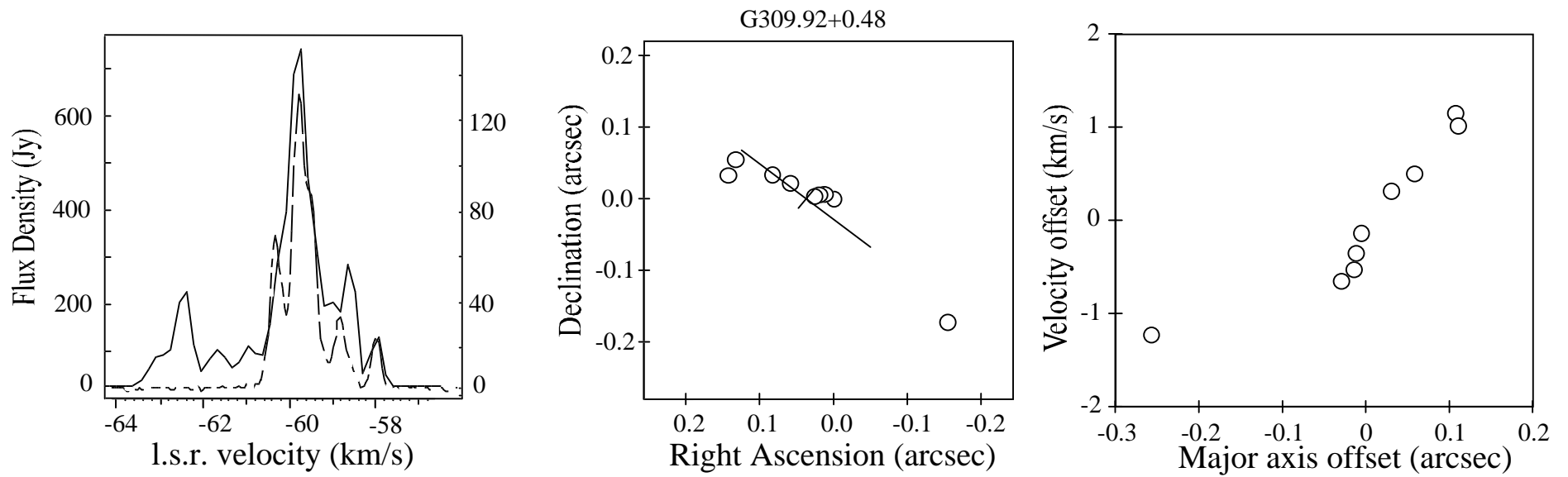

Fig 1. *Single-dish spectrum (a), VLBI map (b), and v-a diagram (c) of G309.92+0.48. Note that not only does the map (b) show a linear structure, but a linear structure is also present in the v-a diagram, which is consistent with an edge-on rotating ring of masers. 12 GHz and 6.7 GHz spectra are represented by broken and full lines respectively. The flux scale to the left of the spectrum refers to the 6.7 GHz data, and that on the right refers to the 12.2 GHz data.. The cross in (b) represents the position angle and origin of the fitted major axis on to which the positions have been projected to obtain the position offsets shown in (c). The origin of both axes in (c) are arbitrary*

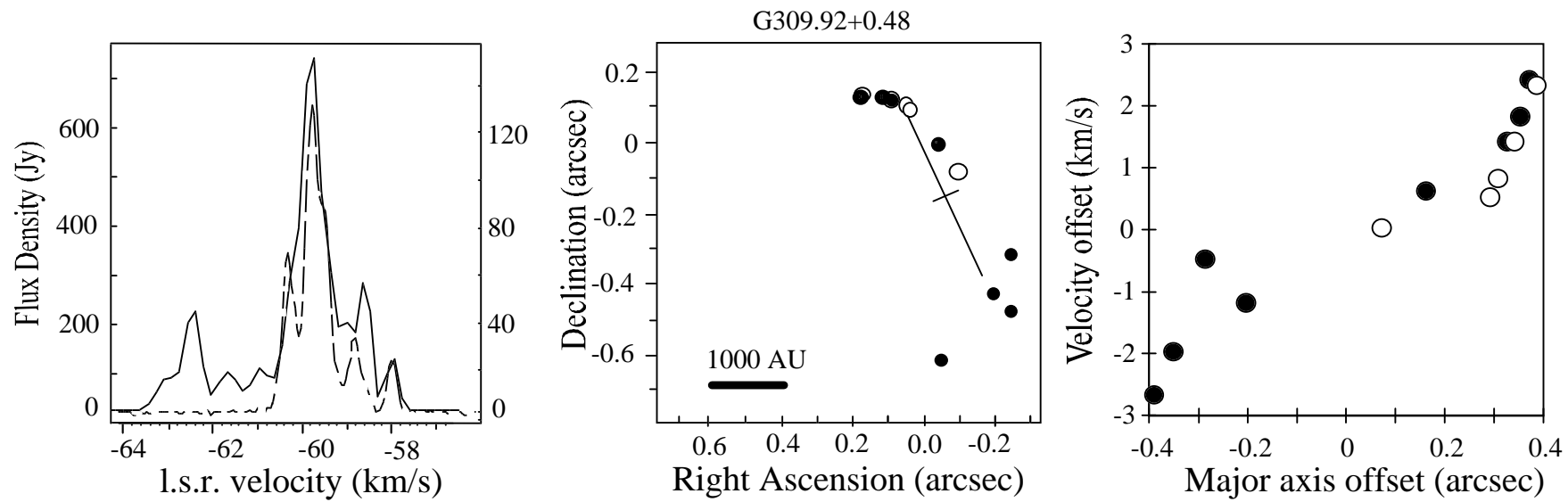

Fig. 2 *Spectrum, map, and v-a diagram of G309.92+0.48. In this and subsequent diagrams: open circles represent 12 GHz masers from N88, and filled circles represent 6.7 GHz masers from N93; 12 and 6.7 GHz spectra are represented by broken and full lines respectively; the flux scale to the left of the spectrum refers to the 6.7 GHz data, and that on the right refers to the 12.2 GHz data; the cross in (b) represents the position angle and origin of the fitted major axis on to which the positions have been projected to obtain the position offsets shown in (c), and the origin of both axes in (c) are arbitrary.*

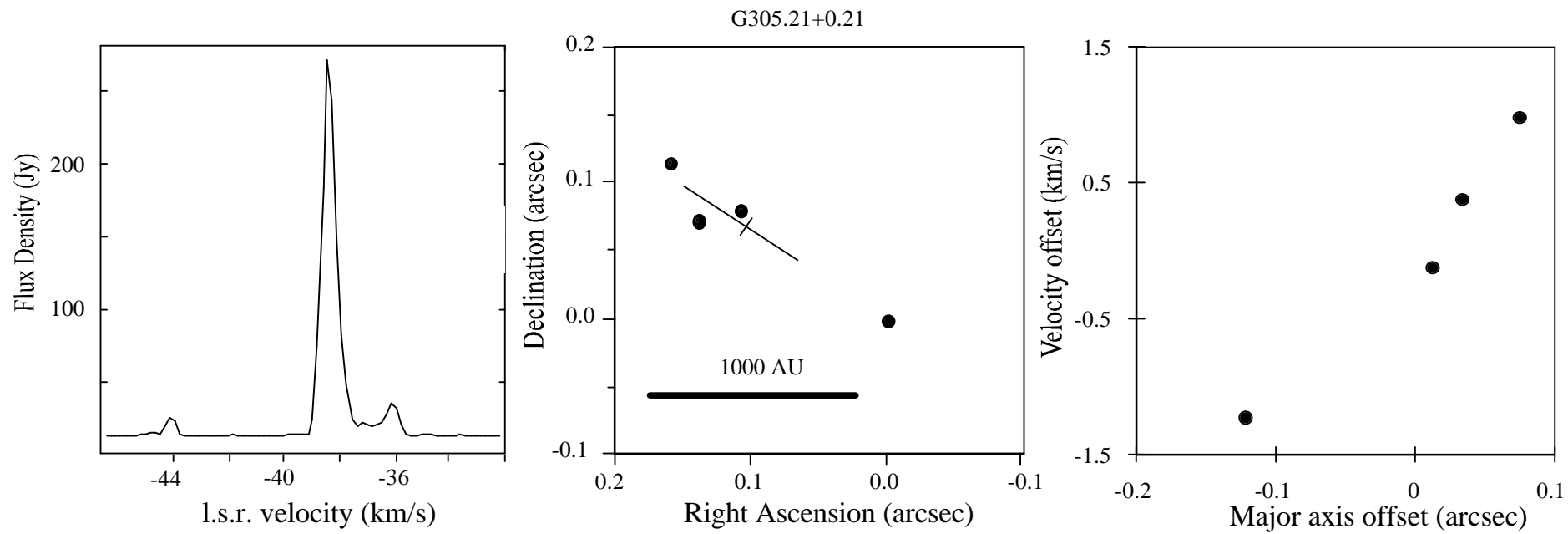

Fig 3. *Spectrum, map, and v-a diagram of G305.21+0.21.*

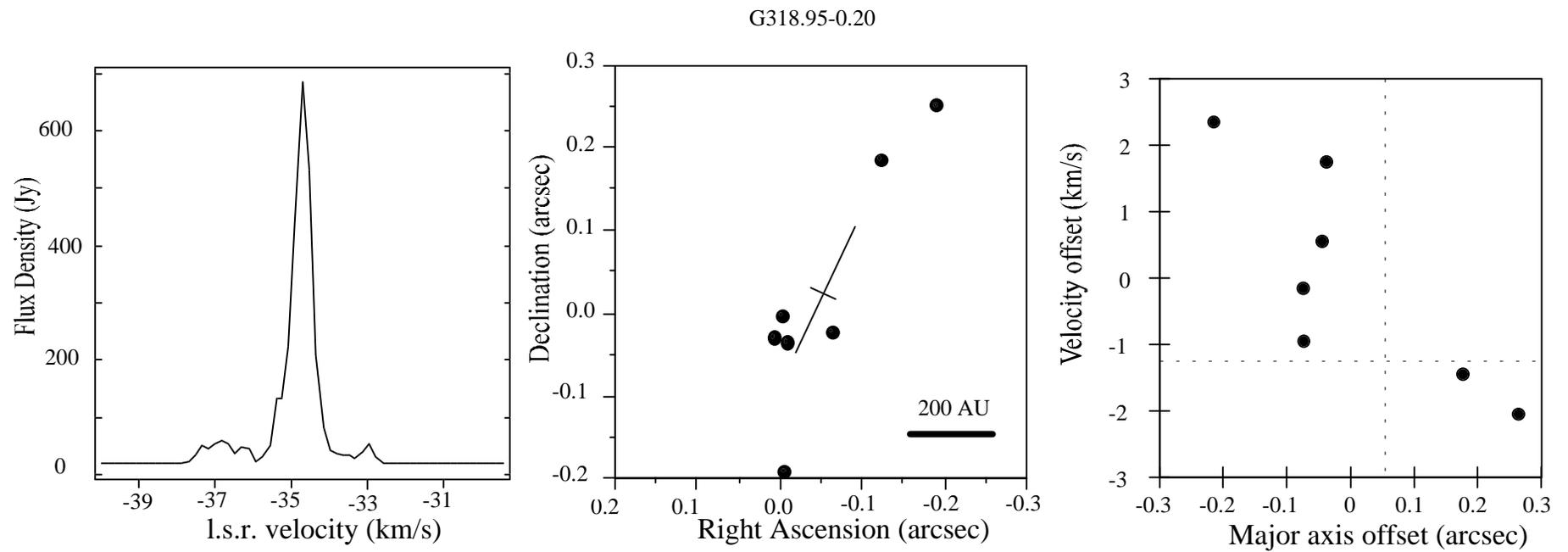

Fig 4. *Spectrum, map, and v-a diagram of G318.95-0.20. The dotted lines in (c), in this and subsequent figures, indicate the position and velocity of the parent star, such that all masers lie in only two quadrants, which is consistent with an edge-on rotating disk.*

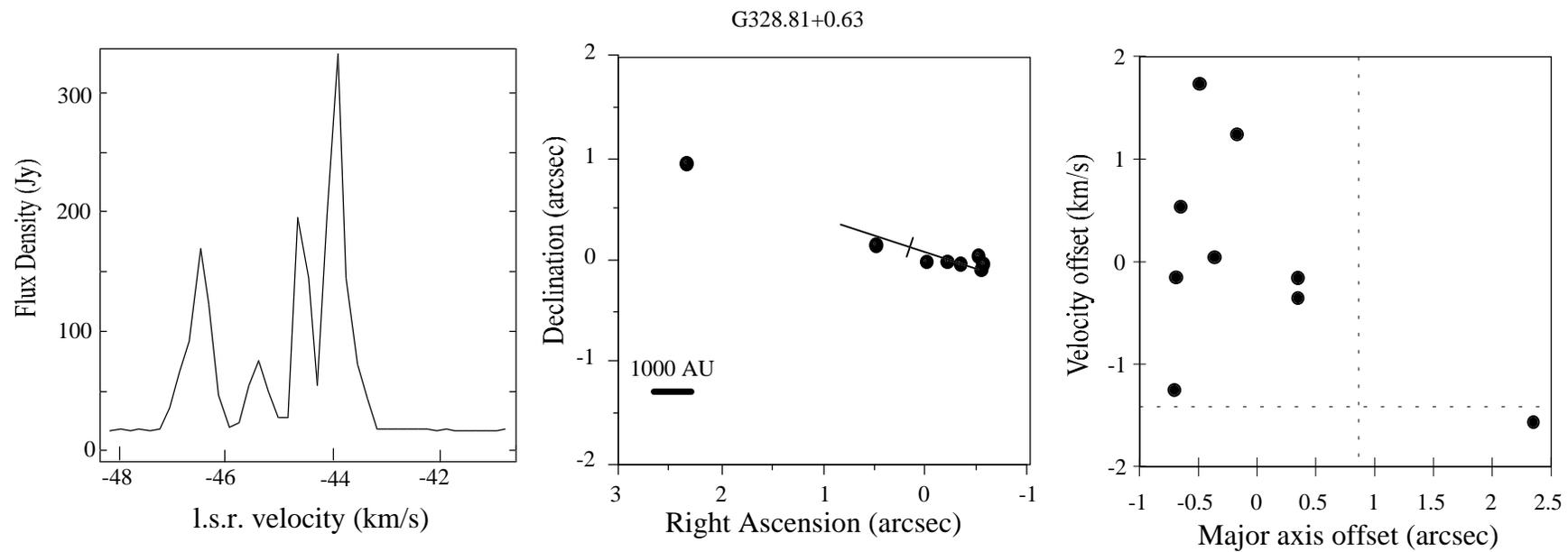

Fig 5. *Spectrum, map, and v-a diagram of G328.81+0.63*

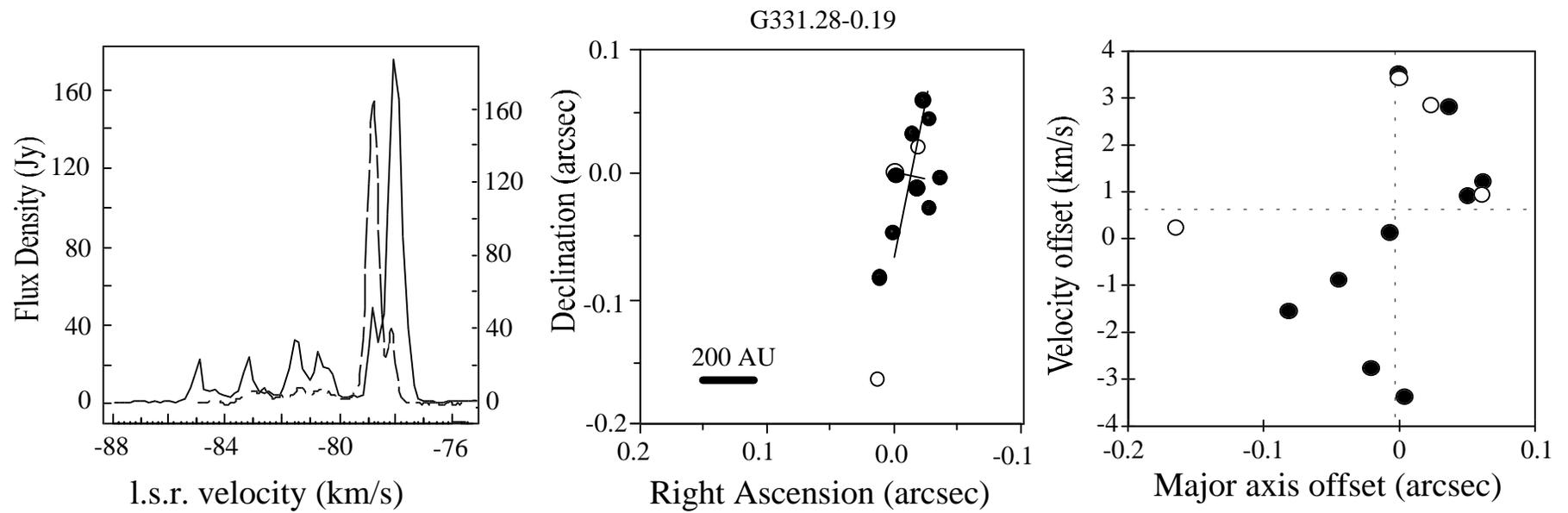

Fig 6. *Spectrum, map, and v-a diagram of G331.28-0.19 The one maser feature in the "wrong" quadrant is consistent with the positional error estimated by N93 of 0.02 arcsec.*

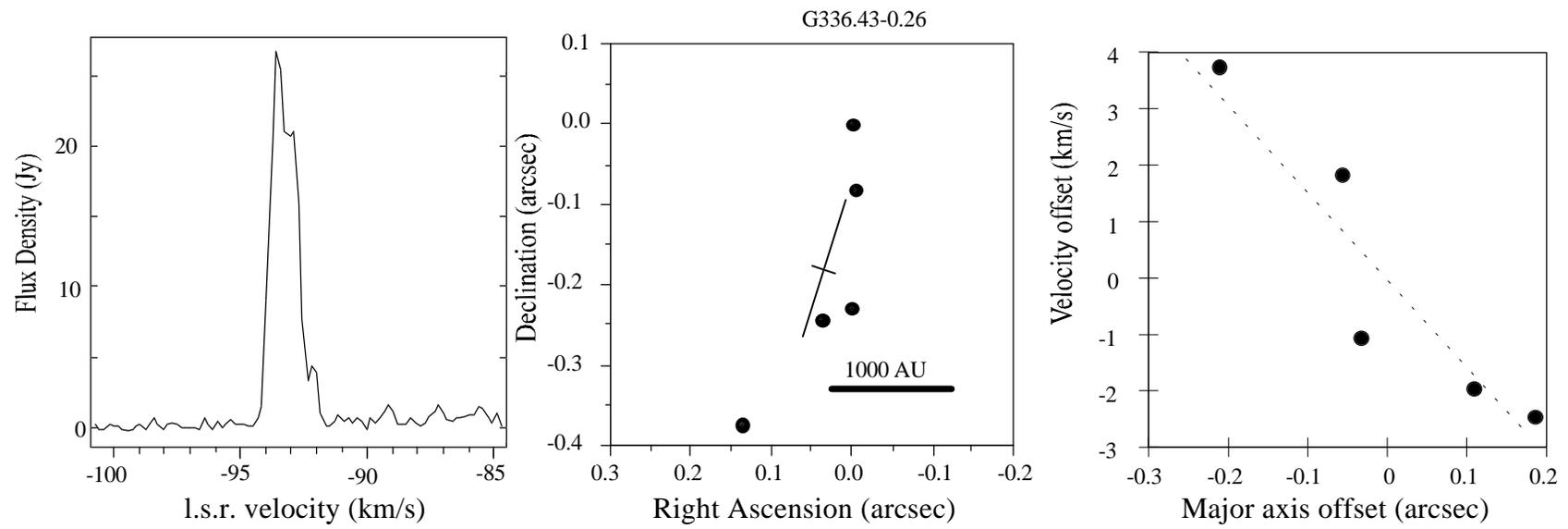

Fig 7. *Spectrum, map, and v-a diagram of G336.43-0.26*

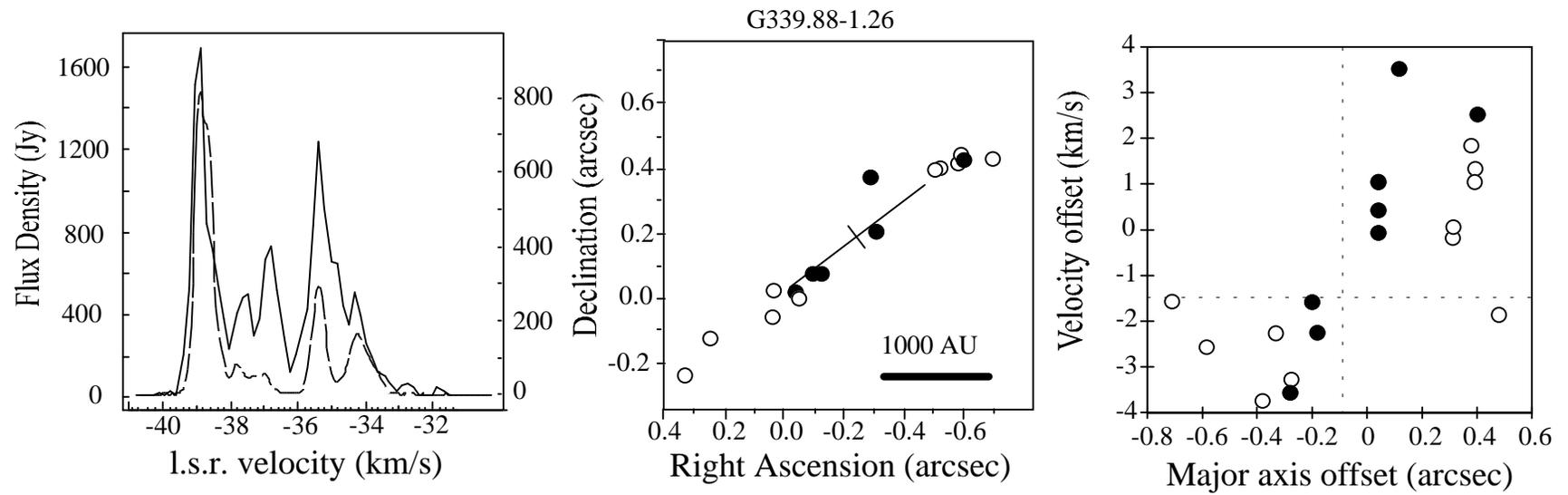

Fig 8. *Spectrum, map, and v-a diagram of G339.88-1.26. The 12 GHz component in the lower right-hand quadrant in (c) (corresponding to "g" in N88 and N93) is anomalous.*

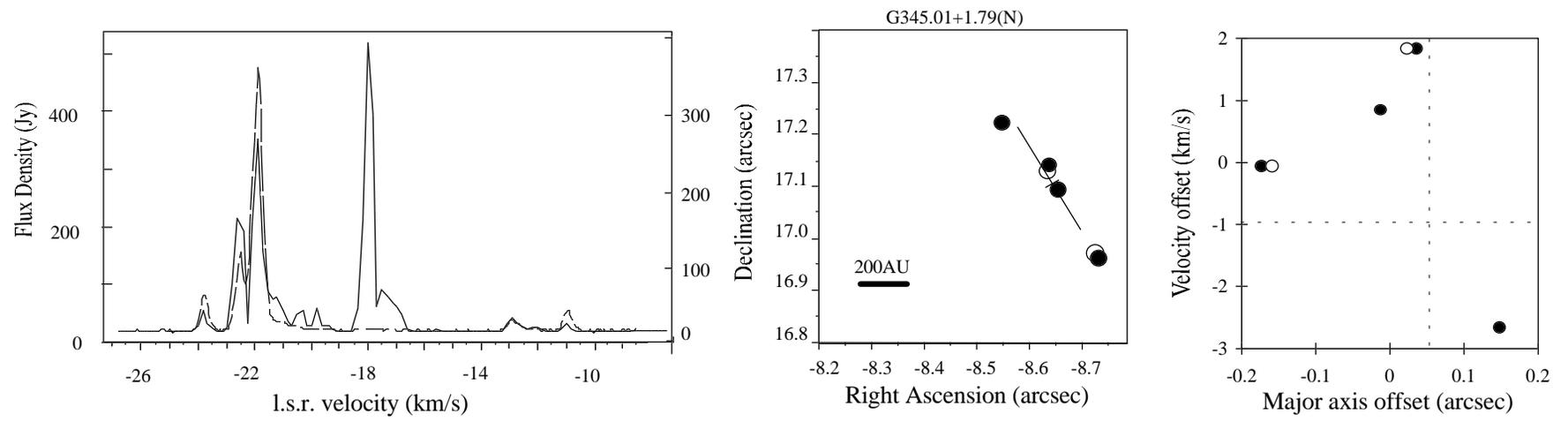

Fig 9. *Spectrum, map, and v-a diagram of G345.01+1.79(N)*

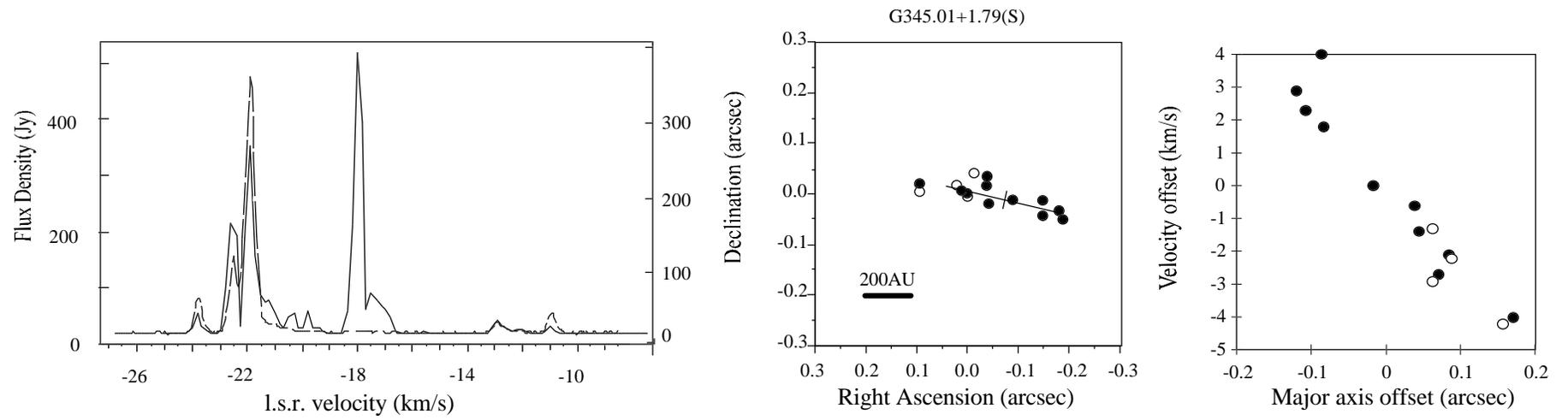

Fig 10. *Spectrum, map, and v-a diagram of G345.01+1.79(S)*

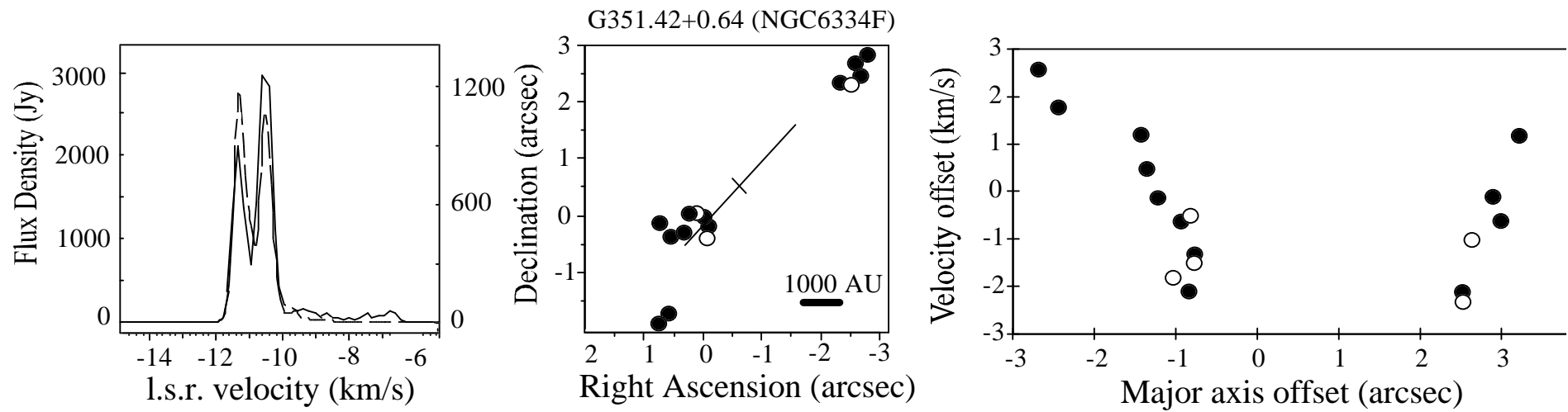

Fig 11. *Spectrum, map, and v-a diagram of G351.42+0.64. The two clusters in the map (b) clearly appear as two lines in the v-a plane, indicating two separate rotating circumstellar disks around separate parent stars, with opposite senses of rotation.*

(a) 3-D view of disk

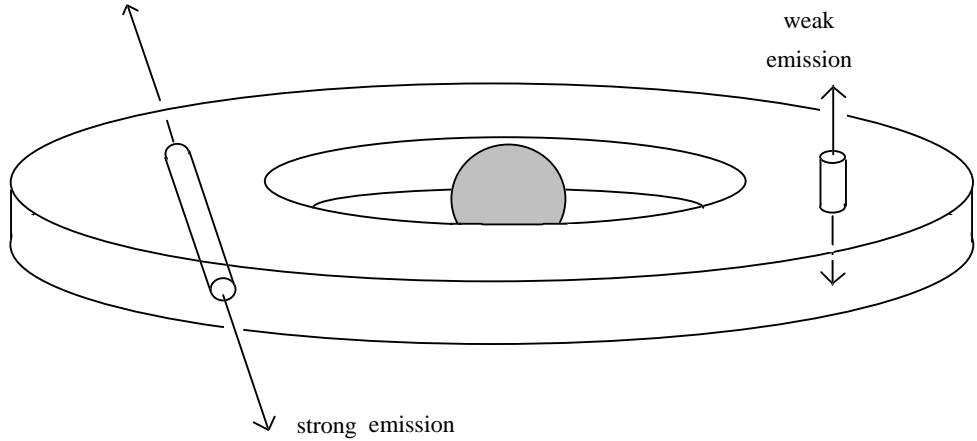

(b) cross-section of disk

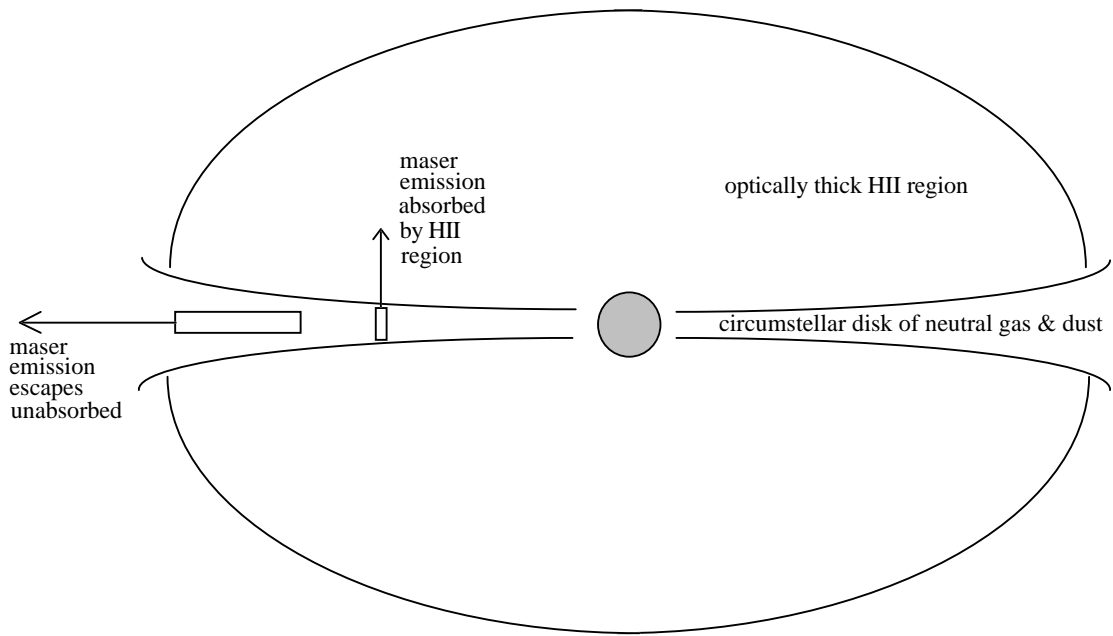

Fig. 12 *Diagram showing the two mechanisms which lead us to view edge-on disks preferentially. In (a) the masers within the plane of the disk have a greater column length than those perpendicular to it, and so are brighter. In (b), any maser emission emitted perpendicular to the disk is absorbed by the hemispherical optically-thick HII regions above and below the plane, whereas masers that emit within the plane suffer no absorption.*

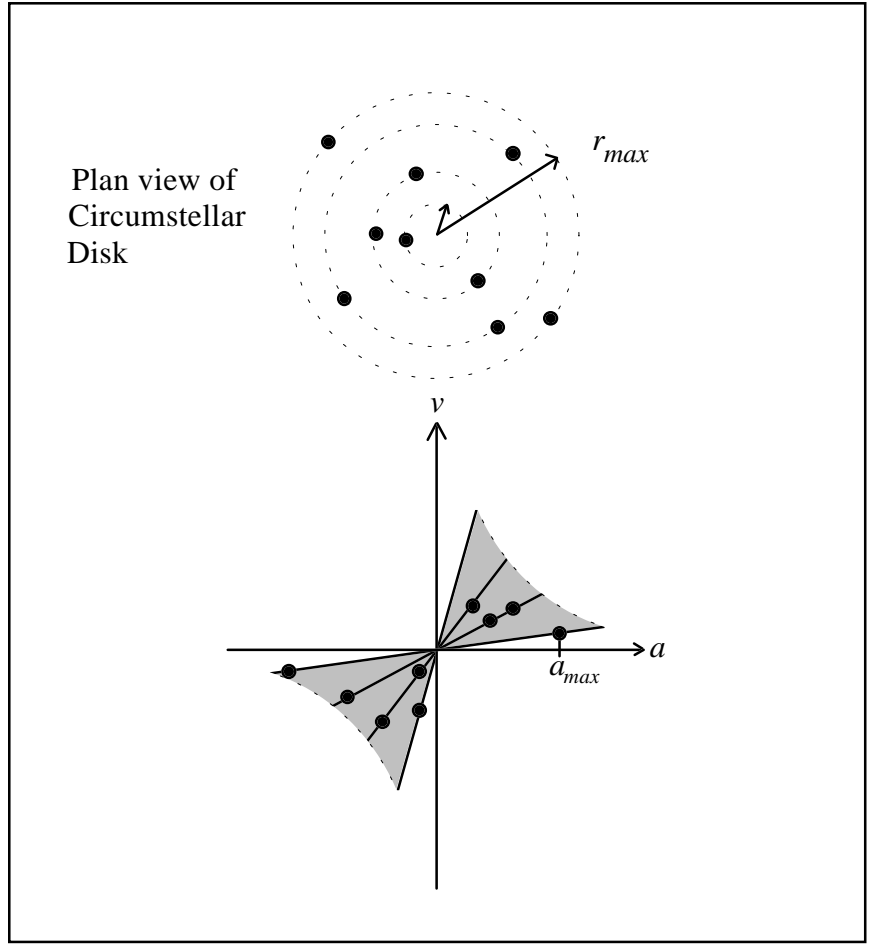

Fig 13. *The rotation curves of a rotating disk. Each broken ring in the upper part of the diagram corresponds to a line in the lower part. The shaded area indicates the locus in the v-a diagram where masers might be found.*